\documentclass[journal]{IEEEtran}

\usepackage{epstopdf}
\usepackage{graphicx}

\usepackage[cmex10]{amsmath}
\DeclareMathOperator\erfc{erfc}

\begin{document}

\title{Gateway Switching in Q/V Band \\ Satellite Feeder Links}
\author{Ahmad~Gharanjik,~\IEEEmembership{Student~Member,~IEEE,}
          Bhavani Shankar Mysore Rama Rao,\\~\IEEEmembership{}Pantelis-Daniel~Arapoglou,~\IEEEmembership{Member,~IEEE,}
          and~Bj\"orn~Ottersten,~\IEEEmembership{Fellow,~IEEE}
\thanks{A. Gharanjik, B. Shankar and B. Ottersten are with the Interdisciplinary Centre for Security, Reliability and Trust (SnT), University of Luxembourg (e-mail: {Ahmad.Gharanjik, Bhavani.Shankar, Bjorn.Ottersten}@uni.lu).
}
\thanks{P.-D. Arapoglou is with ESA/ ESTEC, Noordwijk, Netherlands (e-mail: pantelis-daniel.arapoglou@esa.int).}
\thanks{B. Ottersten is also with the Signal Processing Laboratory, ACCESS Linnaeus Center, KTH Royal Institute of Technology, Sweden.}
\vspace{-0.21in}}
\markboth{}%
{Shell \MakeLowercase{\textit{et al.}}: Bare Demo of IEEEtran.cls for Journals}
\maketitle
\begin{abstract}
A main challenge towards realizing the next generation Terabit/s broadband satellite communications (SatCom) is the limited spectrum available in the Ka band. An attractive  solution is to move the feeder link to the higher Q/V band, where more spectrum is available. When utilizing the Q/V band, due to heavy rain attenuation, gateway diversity is considered a necessity to ensure the required feeder link availability. Although receive site diversity has been studied in the past for SatCom, there is much less maturity in terms of transmit diversity techniques. In this paper, a modified switch and stay combining scheme is proposed for a Q/V band feeder link, but its performance is also evaluated over an end-to-end satellite link. The proposed scheme is pragmatic and has close to optimal performance with notably lower complexity.
\end{abstract}
\IEEEpeerreviewmaketitle
\section{Introduction}
\IEEEPARstart{T}{he} key challenge to achieve a Terabit/s broadband satellite communication (SatCom) system is the limited available spectrum in the currently used Ka-band (20/30 GHz). An attractive solution for resolving the issue is moving the feeder link  from Ka-band to the Q/V-band (40/50 GHz)\footnote{Q/V bands span 35-75 GHz but main frequency allocations for SatCom are the FSS bands, 37.5-42.5 GHz and 47.5-51.4 GHz. }. This migration  provides for higher feeder link bandwidth that can accommodate a broadband SatCom system with a high number of beams ($>$200) and aggressive frequency reuse. Further, it can free-up the whole Ka-band spectrum for the user link. However, heavy fading caused by rain attenuation in Q/V band necessitates the use of gateway (GW) diversity techniques to ensure the required availability \cite{daniel1}.

Although GW site diversity reception is a familiar and mature technique with rich literature \cite{daniel2}, very little attention has been given in SatCom systems on realizing a transmit gateway diversity scheme in the forward link. Equal Gain Combining (EGC) and Maximum Ratio Combining (MRC) have been studied in \cite{daniel1} towards achieving transmit diversity. However, these techniques require accurate channel phase information while both the GWs need to be active, which demands challenging synchronization processes.
Switch and Stay Combining (SSC) and Selection Combining (SC)  which do not require phase information and employ a single active transmitter at any instance have been proposed in terrestrial communications \cite{alo1}-\cite{micha}.
In this paper, building on the SSC, we propose and analyze a novel diversity scheme for Q/V band feeder links suffering from correlated rain fading.
This Modified SSC (MSSC) scheme exploits beacons for attenuation measurement and activates only one GW in a manner that lowers the GW switching rate without performance degradation.
This makes it an ideal candidate for SatCom and avoids frequent GW switching that causes system overhead. Further, MSSC does not warrant any modification of the user terminal and naturally lends itself to the smart GW concepts that have been proposed recently for multi-GW configurations \cite{barry}. Apart from the feeder link, also the benefit of this diversity scheme over the end-to-end (feeder and user) link is analyzed.
\section{System Model}
\label{sec:Sys_Mod}
\begin{figure*}[t]
\hspace{0.2cm}
\begin{tabular}[t]{c}
\begin{minipage}{17 cm}
\begin{eqnarray}
\label{joint_pdf}
f_{A_{1}, A_{2}}\!(A_{1}, A_{2})=\frac{1}{2\pi A_{1} A_{2}\sigma_1\sigma_2\sqrt{1-\rho^2}}
\exp\!{\Bigl(-\frac{1}{1- \rho^2} \Bigl[ u_1^2-2\rho  u_1 u_2+ u_2^{2}\Bigr]\Bigr)}; \; u_i= \frac{\ln{A_i}-m_i}{\sigma_i}, i=1,2
\end{eqnarray}
\end{minipage}
\vspace{0.1cm}\\
\hline
\end{tabular}
\vspace{-0.4cm}
\end{figure*}
Two gateways, $GW_1$ and $GW_2$, separated on ground by a distance of $D$ km communicate with a geostationary satellite over a feeder link operating in the Q/V band with only one of them being active in each transmission time slot. Assume that the active GW transmits the signal $s(t)$ having an average power $E_1=\textbf{E}\{|s(t)|^2\}$. The decision on switching is taken at discrete time instants $t=nT$, where $n$ is an integer and the $T$ is the interval between switching instants. The channel between $GW_i$ and the satellite at $t=nT$ is denoted by $h_i[n]=|h_i[n]|e^{j\alpha_i}, i=1,2$ where $\alpha_i$ is the phase component. The channel amplitude, $|h_i[n]|$, can be estimated using a beacon signal received from the satellite.
The clear sky signal-to-noise ratio (SNR) for the feeder uplink is then defined as  $\gamma_{{CS}_{UL}}=E_1/N_1 $ where $N_1$ is the noise variance at the satellite front-end. The actual SNR for the link between $GW_i$ and the satellite at  $t=nT$ can be obtained by $\gamma_{i, n}= E_1 |h_i[n]|^2/N_1 =|h_i[n]|^2 \gamma_{{CS}_{UL}},\, i=1, 2$.

In the Q/V band the main impairment is the rain attenuation which is typically modeled by the lognormal distribution \cite{p618}.
The other clear-sky effects are assumed to be compensated by a fixed fade margin or an uplink power control scheme. The rain attenuation
and the channel gains are related as $ A_{i, n}=-10\, \log_{10} |h_i[n] |^2, i=1, 2$. The joint probability density function (PDF) of the correlated and identical rain attenuations on the two feeder links, takes the form given in (\ref{joint_pdf}) at the top of next page (for simplicity we drop n). Here, $m_i$ and $\sigma_i$ are the mean and standard deviation of $\ln\negthinspace{A_i}$ respectively. Also, $\rho$ is the spatial correlation coefficient for two $GW$s separated by a distance of $D$ km.
It can be obtained from \cite{itu-p1815} as
$\rho = 0.94\,\exp\left(-{D/30}\right)+0.06\,\exp[-\left({D/500}\right)^2].$

According to the MSSC scheme,  gateway switching from active $GW_i$ to the alternative $GW_j$ occurs if $\gamma_{i, n} <\theta$ and $\gamma_{j, n}> \theta, i\neq j$ where $\theta$ is the switching threshold. In contrast, conventional SSC results in switching when $\gamma_{i, n} <\theta$ regardless of $\gamma_{j, n}$, while in SC, switching always ensures that the active GW has the higher SNR (irrespective of its relation to $\theta$). The proposed MSSC strategy can be implemented without any feedback, with each GW estimating its SNR by employing a beacon signal from the satellite (no phase information needed). In case of switching, the traffic is rerouted to the redundant GW via a terrestrial fibre interconnection.

\section{Performance Analysis}
\subsection{Outage Analysis of the Feeder Link}
\label{ssec:Outage}
We now study the outage performance of MSSC. Denoting the SNR of the active feeder link by $\gamma_n$,  it follows that,
\begin{equation}
\label{gamma_n}
\gamma_n\negthinspace=\negthinspace\gamma_{1,n}\negthinspace\negthinspace\negthinspace \iff\negthinspace\negthinspace\negthinspace\begin{cases} \gamma_{n-1}=\gamma_{1,n-1}\;  , \; \gamma_{1,n}\geq \theta \\
\gamma_{n-1}\negthinspace=\gamma_{1,n-1}\;,
\; \gamma_{1,n} < \theta \; ,\; \gamma_{2,n} < \theta \\
\gamma_{n-1}\negthinspace=\gamma_{2,n-1}\; ,\; \gamma_{2,n} < \theta \;, \; \gamma_{1,n} \geq \theta
\end{cases}
\end{equation}
for the MSSC. Further, $\gamma_n=\gamma_{2,n}$  can be obtained similarly. The cumulative distribution function (CDF) of $\gamma_n$ follows as,
\begin{eqnarray}
\label{gamma_n_cdf}
F_{\gamma_n}\negthinspace\negthinspace(u)\negthinspace\negthinspace=\negthinspace\negthinspace\Pr\{\gamma_{n}\negthinspace\negthinspace=\negthinspace\negthinspace\gamma_{1,n},\gamma_{1,n}\leq
u \}\negthinspace+\negthinspace\Pr\{\gamma_{n}\negthinspace=\negthinspace\gamma_{2,n}\,, \gamma_{2,n}\leq u \}.\negthinspace\negthinspace\negthinspace\negthinspace
\end{eqnarray}
Using (2) and the fact that $\gamma_{1,n}$ and $\gamma_{2,n}$ are identical, (3) can be further simplified by following an approach similar to Appendix of \cite{dayya} as,
\begin{eqnarray}
\label{dayya_simp}
F_{\gamma_n}(u)\negthinspace\negthinspace\negthinspace\negthinspace\negthinspace\negthinspace&=&\negthinspace\negthinspace\negthinspace\negthinspace\negthinspace\negthinspace \Pr\{\theta \leq \gamma_{1,n}\leq u\}+\Pr\{\theta\leq \gamma_{1,n}\leq u \,,\gamma_{2,n}\leq \theta\}\nonumber\\
&&\negthinspace\negthinspace\negthinspace\negthinspace\negthinspace+\Pr\{\gamma_{1,n}\leq\theta\;,\; \gamma_{1,n}\leq u\;,\;\gamma_{2,n}\leq\theta\}\;.\negthinspace\negthinspace
\end{eqnarray}
A system outage occurs if $\gamma_n\negthinspace <\negthinspace \gamma_{th}$, where the outage threshold $\gamma_{th}$ depends on the operational set-up. The outage probability, $ P_{out}(\gamma_{th})\negthinspace\negthinspace = F_{\gamma_n}\negthinspace(\gamma_{th})$, can be obtained from  (\ref{dayya_simp}) as,
\begin{eqnarray}
\label{P_out_feeder}
\negthinspace\negthinspace\negthinspace\negthinspace\negthinspace\negthinspace\negthinspace P_{out}(\gamma_{th})\negthinspace\negthinspace\negthinspace\negthinspace&=&\negthinspace\negthinspace\negthinspace\negthinspace\negthinspace\Pr\{\gamma_{1,n}\leq\theta, \gamma_{1,n}\leq \gamma_{th},\gamma_{2,n}\leq\theta\}+\nonumber\\
&&\negthinspace\negthinspace\negthinspace\negthinspace\negthinspace\negthinspace\negthinspace\negthinspace\negthinspace\negthinspace\negthinspace\negthinspace\negthinspace\negthinspace\negthinspace\negthinspace\negthinspace\negthinspace\negthinspace\negthinspace\negthinspace\negthinspace\negthinspace\negthinspace\negthinspace\negthinspace\negthinspace\negthinspace\negthinspace\negthinspace\negthinspace\negthinspace\negthinspace\negthinspace\negthinspace\negthinspace\Pr\{\theta\leq \gamma_{1,n}\leq \gamma_{th}, \gamma_{2,n}\leq \theta\}+\Pr\{\theta \leq \gamma_{1,n}\leq \gamma_{th}\}\;.
\end{eqnarray}
Setting a predetermined $\theta$ is an important system design issue and significantly affects $P_{out}$ of the system. For a given $\gamma_{th}$, the optimal $\theta$ minimizing $P_{out}$  is given by $\theta\negthinspace\negthinspace=\negthinspace\negthinspace\gamma_{th}$ \cite[Ch.9.8.1]{alo2}.
In this case,  (\ref{P_out_feeder}) reduces to the outage of the SC scheme,
\begin{equation}
\label{P_out_SC}
P_{out}(\gamma_{th}) =\Pr\{\gamma_{1,n}\leq \gamma_{th}\;, \; \gamma_{2,n}\leq
\gamma_{th}\}\;.
\end{equation}
Using the expression of $\gamma_{i,n}$ from Section II in (\ref{P_out_SC}), we have,
\begin{eqnarray}
\label{P_out_final}
P\!_{out}(\gamma_{th})\negthinspace\negthinspace\negthinspace\negthinspace&=&\negthinspace\negthinspace\negthinspace\negthinspace\negthinspace\Pr\{10^{-{A_{1}\over 10}}\gamma_{{CS}_{UL}} \leq \gamma_{th}\;
 ,\; 10^{-{A_{2}\over 10}}\gamma_{{CS}_{UL}} \leq \gamma_{th}\} \nonumber\\
&=&\negthinspace\negthinspace\negthinspace\negthinspace\negthinspace\Pr\{A_{1} > \Gamma_{CS}-\Gamma_{th} \;,\;  A_{2} >\Gamma_{CS}-\Gamma_{th}\},
\end{eqnarray}
where $\Gamma_{CS}\negthinspace\negthinspace=\negthinspace10\log\gamma_{{CS}_{UL}}$ and $\Gamma_{th}=\negthinspace\negthinspace10\log\gamma_{th}$.
The expression for the outage probability can be derived simply as
\begin{equation}
\label{p_out_int}
 P_{out}(\gamma_{th})\negthinspace=\negthinspace\!\int_{\Gamma_{CS}-\Gamma_{th}}^\infty \int_{\Gamma_{CS}-\Gamma_{th}}^\infty \negthinspace\negthinspace\negthinspace\negthinspace{f_{{A_1},A_2}\!\left(A_{1},A_{2} \right)}\;d{A_{1}}d{A_{2}}\,.
\end{equation}
Using \cite[Eq. 233.1.8]{grad} and after some manipulation, $P_{out}$ of the MSSC scheme in the feeder uplink can be obtained as,
\begin{equation}
\label{p_out}
P_{out}^{UL}(\gamma_{th})=\frac{1}{2\sqrt{2\pi}} \int _{\beta_2}^{\infty}{\exp\!{\left(\frac{-x^2}{2}\right)}\erfc\!{\left(\frac{\beta_1-\rho
x}{\sqrt{2\left(1-\rho^2\right)}}
 \right)}}\, dx,
\end{equation}
where $\beta_i=\frac{\ln{\left(\Gamma_{CS}-\Gamma_{th}\right)}-m_i}{\sigma_i},  i=1, 2$. This integral can be evaluated numerically.
\vspace*{-0.3in}
\subsection{End-to-End Outage Analysis}
Vast majority of SatCom systems are transparent $-$ the satellite repeater only downconverts the signal received on the feeder link and amplifies it  before re-transmitting onto the user link. Given that the user link will operate in a  band (like Ka) lower than the feeder link, it is interesting to investigate the improvement of the end-to-end link due to MSSC. Although a similar geometry has been modelled in terrestrial dual hop radio relay system \cite{sakar}, to the best of our knowledge this is first time the satellite link is analyzed for this diversity technique.

Towards this, the satellite repeater gain, denoted by $G_s$, ensures that the output power level is fixed to $E_2$. Therefore, the amplifying factor can be obtained by
\begin{equation}
\label{Amp_factor}
G_s^2={E_2}/(|h[n]|^2E_1+N_1)
\end{equation}
where $h[n]$ is the corresponding channel of the active GW. The signal received by the user terminal from the satellite is $r'[n]=g[n]G_s(h[n] s[n]+n_1[n])+n_2[n]$ where $g[n]$ is the channel between the satellite and the user, while  $n_2[n]$ is the receiver additive white Gaussian noise (AWGN) of variance $N_2$. The equivalent SNR at the receiver can be written as,
\begin{equation}
\label{eqv_snr}
\gamma_{eq} = {\gamma_{g}\gamma_{h}}/(\gamma_{g}+\gamma_{h}+1)\;,
\end{equation}
where $\gamma_h=\gamma_n= E_1 |h|^2/N_1$ and $\gamma_g= E_2 |g|^2/N_2$  with the time index $n$ dropped for simplicity. The clear sky SNR for downlink is defined as $\gamma_{{CS}_{DL}}= E_2 /N_2$. Finally, the end-to-end outage probability, $P_{out}^{e2e}(\gamma_{th})$, can be calculated as (\ref{p_out_e2e}) found at the top of the next page, where $f_{\gamma_g}\!(\gamma_g)$ is the PDF of the $\gamma_g$ and $P_{out}^{DL}=\Pr\{\gamma_g<\gamma_{th}\}$. This equation shows the impact of feeder link improvement on the overall performance of the system. It is worth mentioning that the lower bound (last inequality of (\ref{p_out_e2e}) ) is the outage performance when the satellite is operating in the regenerative mode.
\begin{figure*}[t]
\hspace{0.0cm}
\begin{tabular}[t]{c}
\begin{minipage}{17.5 cm}
\begin{eqnarray}
\label{p_out_e2e}
\negthinspace P_{out}^{e2e}(\gamma_{th})&=&\Pr\!\{\gamma_{eq} \leq \gamma_{th}\}=
\int_{0}^{\infty}{\Pr\!\left(\frac {\gamma_{g}\gamma_{h}}{\gamma_{g}+\gamma_{h}+1} \leq \gamma_{th}|\gamma_g\right)f_{\gamma_g}\!(\gamma_g)}\,d\gamma_g\nonumber\\
&=&\int_{0}^{\gamma_{th}}{\Pr\!\left(\gamma_h >z |\gamma_g\right)f_{\gamma_g}\!(\gamma_g)}\,d\gamma_g+\int_{\gamma_{th}}^{\infty}{\Pr\!\left(\gamma_h \leq z|\gamma_g\right)f_{\gamma_g}\!(\gamma_g)}\,d\gamma_g \; ;\;z=\frac{\gamma_{th}(\gamma_g+1)}{\gamma_g-\gamma_{th}}\nonumber\\
&=&P_{out}^{UL}(\gamma_{th})+\int_{\gamma_{th}}^{\infty}{P_{out}^{UL}\!(z)f_{\gamma_g}\!(\gamma_g)}d\gamma_g>P_{out}^{DL}(1-P_{out}^{UL})
+P_{out}^{UL}\,\;
\end{eqnarray}
\end{minipage}
\vspace{0.02cm}\\
\hline
\end{tabular}
\vspace*{-0.3cm}
\end{figure*}
\vspace*{-0.1in}
\subsection{Switching Rate }
When a GW switching strategy is used in the transmission side, the switching rate becomes an important issue.
Clearly, reduced switching rate for a given performance is desirable from a system implementation and operation
view while a high switching rate can make the system unstable. Towards this, we analyze the switching rate of MSSC
by employing a Markov chain model \cite{alo1}. We define six states as in Table \ref{markov_chain}.

Clearly, whenever the active GW is in state 3 or 6, switching occurs. So, the probability of switching is given by $\pi_{3}+\pi_{6}$ where $\pi_i$ is the probability that GW is in state $i$. Based on the MSSC switching strategy, the transitional probability matrix $\bf P$ of the Markov chain can be obtained as,
\begin{equation}
\bf P = \left(\begin{IEEEeqnarraybox*}[][c]{,c/c/c/c/c/c,}
1-p&p_{12}&p-p_{12}&0&0&0\\
1-p&p_{12}&p-p_{12}&0&0&0\\
0&0&0&1-p&p_{12}&p-p_{12}\\
0&0&0&1-p&p_{12}&p-p_{12}\\
0&0&0&1-p&p_{12}&p-p_{12}\\
1-p&p_{12}&p-p_{12}&0&0&0\\%
\end{IEEEeqnarraybox*}\right).
\end{equation}
\begin{table}[!t]
\renewcommand{\arraystretch}{1.1}
\caption{Markov Chain Modelling}
\label{markov_chain}
\centering
\begin{tabular}{cccccc}
\hline
State&$\gamma_{n}$&$\gamma_{n-1}$ &$\gamma_{1, n}$&$\gamma_{2, n}$&Description \\
\hline
1 &$\gamma_{1,n}$&$\gamma_{1,n-1}$&$\geq\theta$ & $\cdot$ & $GW_1$ continues to be active\\
2 &$\gamma_{1,n}$&$\gamma_{1,n-1}$&$<\theta$ & $\leq\theta$ & 2  GWs in outage, no switching\\
3 &$\gamma_{2,n}$&$\gamma_{1,n-1}$&$<\theta$ & $>\theta$ & $GW_1$ in outage, $GW_2$ better \\
4 &$\gamma_{2,n}$&$\gamma_{2,n-1}$&$\cdot$  &$\geq\theta$& $GW_2$ continues to be active\\
5 &$\gamma_{2,n}$&$\gamma_{2,n-1}$& $\leq \theta$ & $<\theta$ & 2  GWs in outage, no switching\\
6 &$\gamma_{1,n}$&$\gamma_{2,n-1}$& $>\theta$ & $<\theta$ & $GW_2$ in outage, $GW_1$ better\\ \hline
\end{tabular}
\end{table}
Here $p=\Pr\{\gamma_{1,n}\leq \theta\}=\Pr\{\gamma_{2,n}\leq \theta\}$ and $p_{12}=\Pr\{\gamma_{1,n}\leq \theta , \gamma_{2,n}\leq \theta\}$. By using the facts that $\overrightarrow
{\pi}= \overrightarrow{\pi}\bf P$ and $\sum_{i=1}^{6}{\pi_i}=1$, where $\overrightarrow{\pi}=[\pi_1, \pi_2, ..., \pi_6]$, the MSSC switching probability can be calculated as
\begin{equation}
\label{eqv_snr}
P_{sw} =\pi_3+\pi_6=p-p_{12}.
\end{equation}
Finally, the switching rate is calculated by $R_{sw}= (p-p_{12})/T$. The switching probability of the conventional SSC was obtained in \cite{alo1} as $P_{sw}=p$ and for SC easily can be found as 0.5. In Section \ref{ssec:Outage} and in \cite{micha},respectively, it was shown that by selection of a proper switching threshold, both, MSSC and SSC will have the same outage performance as SC. However, (\ref{eqv_snr}) shows MSSC has the advantage of a lower switching rate  compared to both SC and SSC.
\begin{figure}[!t]
\centering
\includegraphics[width=3.3in,height=2.3in]{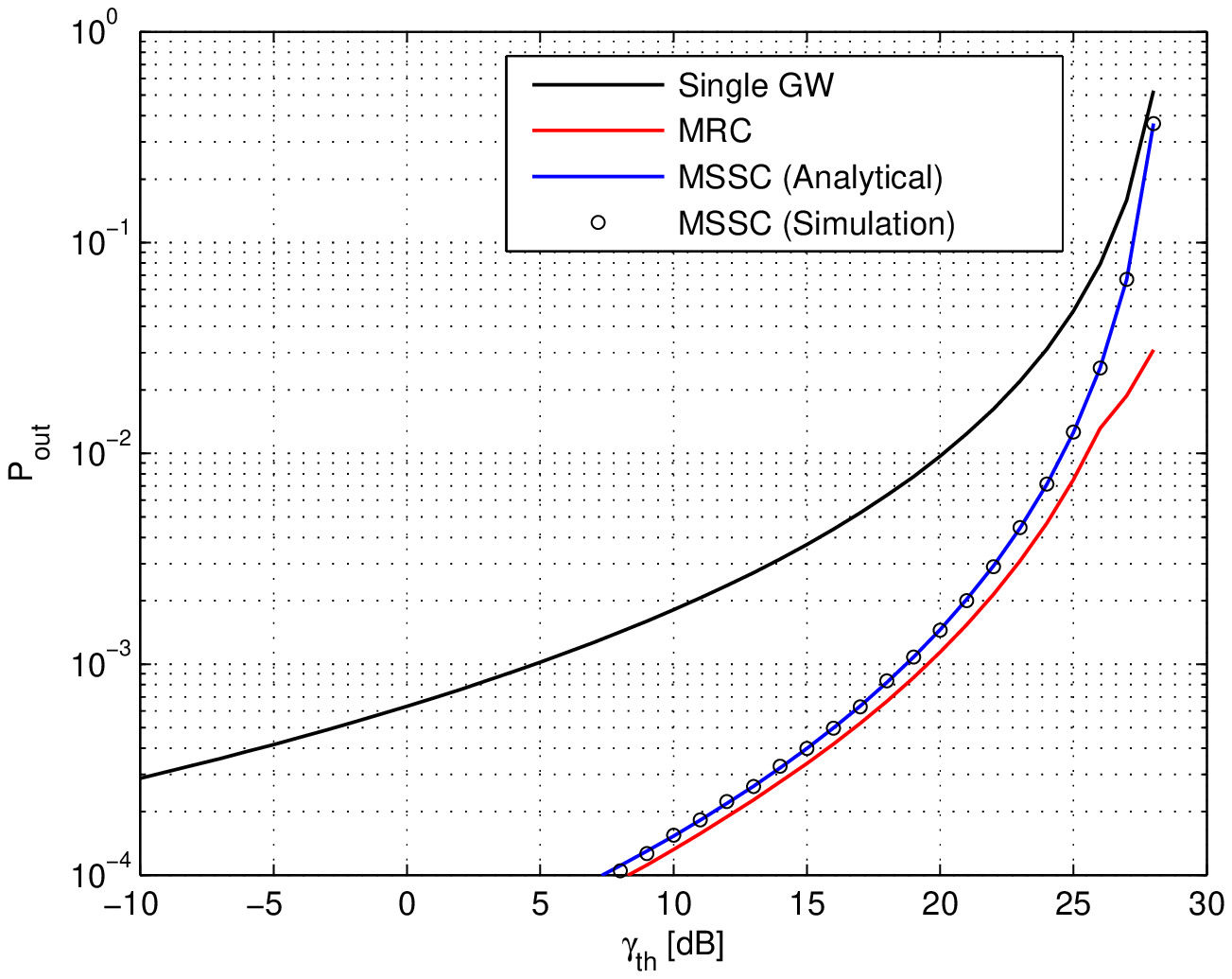}
\caption{Outage of GW diversity strategies on the feeder uplink
($D$=20 Km)}
\label{fig1}
\centering
\includegraphics[width=3.3in,height=2.3in]{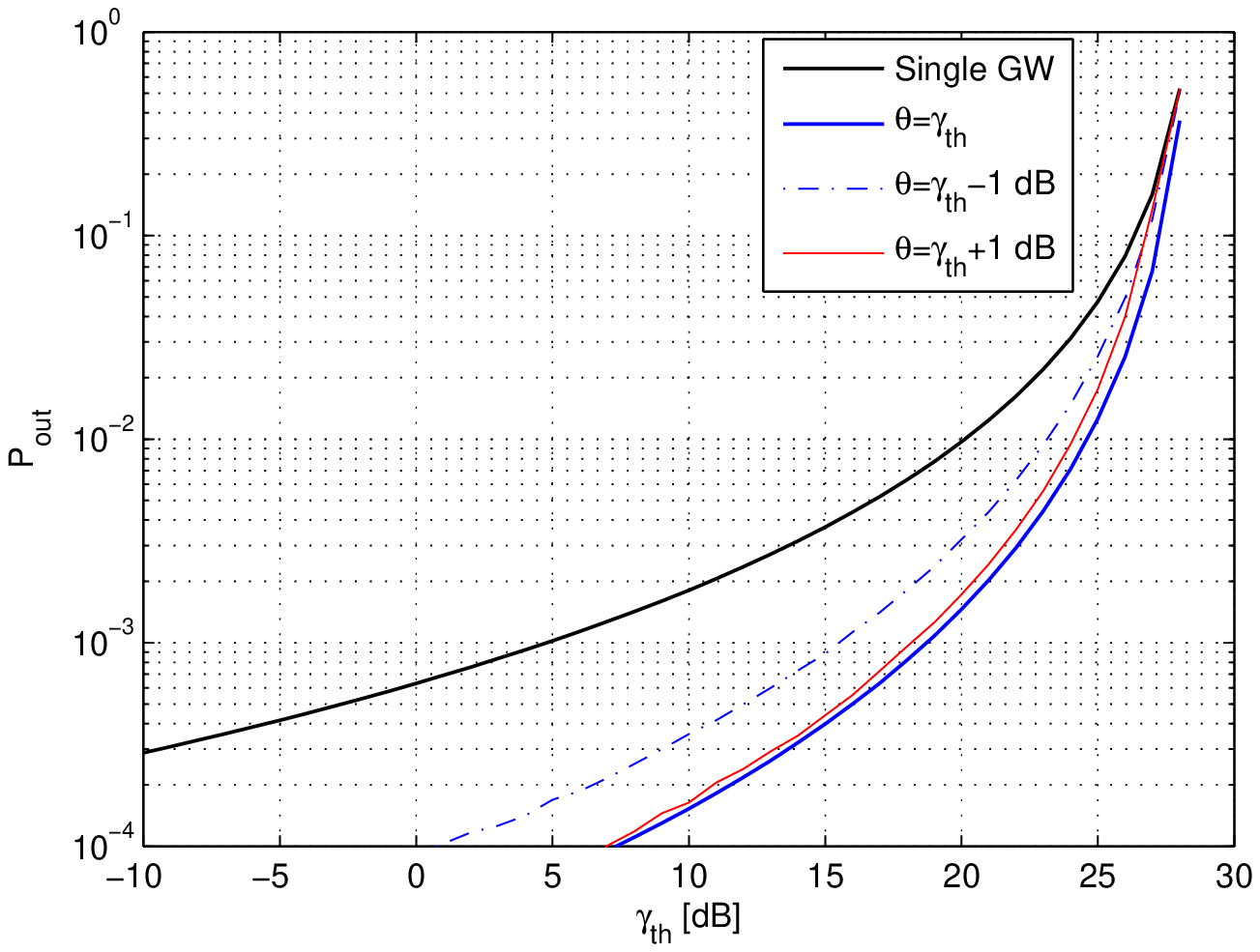}
\caption{Outage results with non-optimal switching threshold $\theta$ ($D$=20
Km)}
\label{fig2}
\end{figure}
\begin{figure}[!t]
\centering
\includegraphics[width=3.3in,height=2.3in]{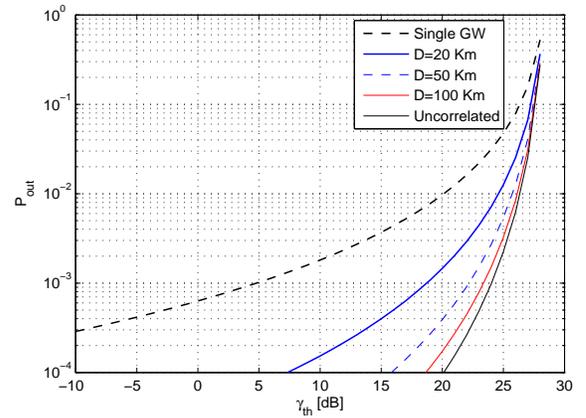}
\caption{Influence of the spatial correlation on the outage performance}
\label{fig3}
\end{figure}
\begin{figure}[!t]
\centering
\includegraphics[width=3.3in,height=2.3in]{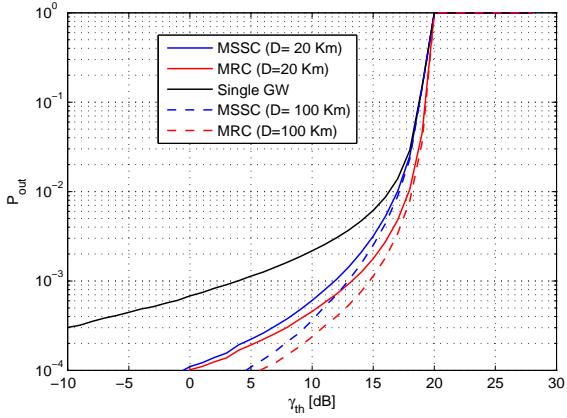}
\caption{End-to-End outage performance of the satellite forward link}
\label{fig4}
\end{figure}
\begin{figure}[!t]
\centering
\includegraphics[width=3.3in,height=2.3in]{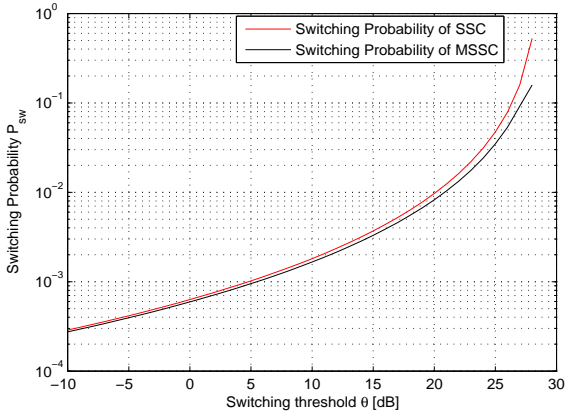}
\caption{Switching Probability of different strategies ($D$=20 K)}
\label{fig5}
\end{figure}
\section{Numerical Results and Discussion}
Table \ref{prop_assump} details the propagation parameters that were used as input to the empirical rain attenuation
prediction model included in ITU-R Recommendation P.618 \cite{p618}. Table \ref{ka_DL} presents the forward-link budget
that has been used in the numerical results.
\begin{table}[!t]
\renewcommand{\arraystretch}{1.3}
\caption{propagation assumptions }
\label{prop_assump}
\centering
\begin{tabular}{c|c}
\hline
\bfseries V Band Up-Link & \bfseries Value\\
\hline
GWs Location & Luxembourg (49.36$^\circ$N;\,6.09$^\circ$E) \\
Carrier frequency & 50 GHz\\
Elevation angle & 32$^\circ$\\
Polarization & Circular\\
\hline
\bfseries Ka Band Down-Link & \bfseries \\
\hline
Receiver Location & Amsterdam (52.3$^\circ$N;\,4.8$^\circ$E) \\
Carrier frequency & 20 GHz\\
Elevation angle & 35$^\circ$\\
Polarization & Circular\\
\hline
\end{tabular}
\end{table}
\begin{table}[!t]
\renewcommand{\arraystretch}{1.3}
\caption{Forward (up/down) link budget}
\label{ka_DL}
\centering
\begin{tabular}{c|c}
\hline
\bfseries Description & \bfseries Value\\
\hline
EIRP$_{GW}$ including back-off & 76.5 dBW\\
UL free space loss & 218.3 dB\\
(G/T)$_{Sat}$ & 31.45 dB\\
$\gamma_{{CS}_{UL}}$ & 28.3 dB\\
EIRP$_{sat}$ including back-off & 72.5 dBW\\
DL free space loss & 210.5 dB\\
(G/T)$_{UT}$ & 20.3 dB\\
$\gamma_{{CS}_{DL}}$ & 21.3 dB\\
\hline
\end{tabular}
\end{table}

Fig. \ref{fig1} compares the analytically obtained outage performance of the proposed scheme with that of MRC \cite{daniel1} on the feeder uplink for GW separation of 20 Km. Also, the Monte-Carlo simulation of the MSSC scheme is plotted to corroborate the analytical results. While these schemes have relatively similar outage performance, MRC is not a realistic option for realizing GW diversity since it assumes that two GWs transmit to the satellite in a synchronized fashion. However, MSSC is not beset with these issues.

Fig. \ref{fig2} illustrates the outage probability of the MSSC for different non-optimal values of the switching threshold ($\theta$). It is clear that the system has the best performance when the switching threshold is set to the outage threshold ($\theta=\gamma_{th}$). It is also worth mentioning that, in the event of an erroneous threshold selection, over-estimation  of $\theta$ yields better outage.

Fig. \ref{fig3}  shows the influence of spatial correlation on the feeder uplink performance. It can be inferred from the plots that for $D>100$ Km, the GWs can be assumed to be spatially uncorrelated.

Fig. \ref{fig4}  plots the end to end outage performance of the system. For a typical availability of 99.9\% (outage $10^{-3}$) diversity gains for MSSC is 7.8 dB  and for MRC is about 9.3 dB compared to single GW when $D=20$ Km. For $D=100$ Km, these values increase to 9 dB and 10.7 dB respectively.

Fig. \ref{fig5} depicts the switching probability of the traditional SSC and the proposed MSSC. It can be seen that the switching probability of the GWs is slightly  improved but at the expense of requiring both GW's SNR unlike the SSC which requires only SNR of the active GW. However, it is not the case for SatCom as the SNR can be easily obtained employing beacon signals.

\section{Conclusion}
In this paper, a modified switch and stay scheme for Q/V band feeder
link has been studied. Although being one of the few realistic GW diversity strategies $-$ since it involves a single GW transmitting at each instant $-$ it has not been hitherto studied for a correlated rain fading channel. Apart from the theoretical analysis of the outage performance, we also address practical issues such as performance of the end-to-end (transparent) link, the effect of erroneous threshold selection, as well as the switching rate between the GWs. Proposed scheme achieves performance comparable to the optimal
one with a lower complexity.

\ifCLASSOPTIONcaptionsoff
  \newpage
\fi

\end{document}